\documentclass[conference]{IEEEtran}
\usepackage{amsmath,amssymb,amsfonts}
\usepackage{graphicx}
\usepackage{float}
\usepackage{tabularx}
\usepackage{booktabs}
\usepackage[table,xcdraw]{xcolor}
\usepackage{makecell}
\usepackage[caption=false,font=footnotesize]{subfig}
\usepackage{cite}
\usepackage{pdfpages}
\usepackage{physics}
\usepackage{listings}
\graphicspath{{images/}{input/images/}{images/Analysis/}}

\begin{document}
\title{AC-OPF Feasibility Analysis and Sensitivity-Guided Capacitor Placement in a High-PV Islanded Microgrid}

\author{\IEEEauthorblockN{1\textsuperscript{st} Aaron Jones}
\IEEEauthorblockA{\textit{Dept. of Electrical Engineering and Computer Science} \\
\textit{Massachusetts Institute of Technology}\\
Cambridge, Massachusetts}
\and
\IEEEauthorblockN{2\textsuperscript{nd} Marija Ilic}
\IEEEauthorblockA{\textit{Dept. of Electrical Engineering and Computer Science} \\
\textit{Massachusetts Institute of Technology}\\
Cambridge, Massachusetts}
}
\maketitle




\begin{abstract}
This paper presents a comparative AC Optimal Power Flow (AC-OPF) study on a real-world city-scale islanded microgrid with high solar PV penetration, implemented within a Digital Twin (DT) framework. Four objective-function cases---economic dispatch, voltage-stress exposure via PV power-factor variation, then optimal load delivery (OLD), and capacitor-enhanced economic dispatch as recovery options are evaluated over a 47-hour time-series horizon on the same network under a shared loading scenario. Optimization sensitivities $\mathrm{OS}_Q$ and $\mathrm{OS}_V$ extracted from all cases are combined into a composite placement score used to rank candidate buses for shunt capacitor upgrades. A post-processing planning optimization balances capacitor-upgrade cost against avoided value-of-lost-load (VoLL), enabling direct economic comparison of infrastructure investment versus reliability penalties. Results demonstrate that sensitivity-guided capacitor placement restores full load service across the horizon and provides targeted reactive support at a quantifiable cost trade-off against corrective load shedding.
\end{abstract}

\section{Introduction}
\label{sec:intro}
Microgrid planning under high distributed energy resource (DER) penetration requires technically defensible methods for assessing how much PV capacity can be integrated---the hosting capacity (HC)---and where corrective voltage-support infrastructure should be placed \cite{ieee1547}. Recent reviews confirm that HC estimation remains an active area, with methods spanning deterministic power-flow studies, stochastic Monte Carlo analysis, and optimization-based approaches \cite{mousa2024hcreview}. While many of these methods target grid-connected distribution feeders, fewer address the distinct challenge of \emph{islanded} microgrids where limited external voltage support is available.
In islanded operation, voltage feasibility is governed entirely by local generation and reactive resources. AC-OPF retains the full nonlinear power-flow equations along with device and voltage constraints, making it the appropriate steady-state tool when voltage magnitude and reactive-power interactions are significant \cite{zimmerman2011matpower}. Linearized (DC-OPF) methods that omit these quantities can understate the feasibility boundary at high PV penetration \cite{bolognani2016existence}. Prior work has established that exact AC power-flow satisfaction is necessary for credible islanded microgrid operations and that voltage collapse risk in islanded microgrids is structurally tied to reactive support adequacy \cite{eajal2018existence}.
The AC Extended Optimal Power Flow (AC XOPF) framework introduced by Ili\'{c} et al.\ supports multiple performance criteria within a unified software tool: elastic feasibility, economic efficiency, and minimization of load not served \cite{cvijic2018xopf}. Optimization sensitivities generated by the XOPF are used to identify the most effective scheduling and, when necessary, the critical infrastructure upgrades. The DyMonDS Digital Twin platform provides the multi-layered simulation environment in which these solves are embedded \cite{ilic2025dymonds}. Recent extensions have applied this framework to probabilistic deliverability assessment of DERs at bulk-power scale \cite{anton2025pda}. The present paper is the first to deploy the four-case AC XOPF methodology at the islanded-microgrid scale for sensitivity-guided capacitor planning.
Microgrids are also recognized as key assets for power-system resilience, and their power-flow under DER variability has been surveyed extensively. Corrective control using microgrids as flexible resources under uncertainty has been demonstrated through robust security-constrained OPF formulations \cite{zhang2017robustSCOPF}. The present study complements these threads by providing a workflow that bridges \emph{operational} AC-OPF analysis to \emph{planning} decisions via optimization sensitivities. The specific contributions of this paper are:
\begin{enumerate}
  \item A four-case comparative AC-OPF study on a real-world islanded microgrid under a shared high-PV scenario, showing how Economic Baseline, Voltage-Stress, Optimal Load Delivery(OLD), and Cap-Enhancement simulations produce complementary planning signals.
  \item A composite sensitivity score $S_k$ combining reactive ($\mathrm{OS}_Q$) and voltage ($\mathrm{OS}_V$) sensitivities to rank capacitor placement candidates, with adjustable weights.
  \item A post-processing planning optimization that balances capacitor investment against avoided VoLL, enabling direct economic comparison of the capacitor-enhanced case against corrective load shedding.
  \item Demonstration that sensitivity-guided capacitor placement restores time-series feasibility across the full horizon, with the specific placement depending on both grid topology and the objective formulation used---highlighting the planning-relevant information gained from the four-case comparison.
\end{enumerate}

\section{Background and Study Context}
\label{sec:background}
\subsection{Voltage Feasibility in Islanded Microgrids}
Voltage instability arises when the nonlinear AC power-flow equations lose admissible solutions under given loading, generation, and reactive-support conditions \cite{kundur2004definition}. In large interconnected systems this behavior is associated with a saddle-node bifurcation of the power-flow equations \cite{canizares2002}. In islanded microgrids with high PV penetration, a structurally similar phenomenon emerges: real power may be sufficient in aggregate, but insufficient or uncoordinated reactive support prevents the network from sustaining voltages within prescribed bounds. This separates the operational limit from generation capacity---the system can become infeasible even when active-power energy is available because no admissible steady-state operating point exists.
In this work, AC-OPF infeasibility is treated as operational evidence that the constrained nonlinear equations admit no solution for a given time step, network state, and demand level. These results show that solution existence depends jointly on loading, topology, and reactive support \cite{bolognani2016existence}. The digital-twin framework enables repeatable time-series scenario analysis of the given system under various optimization objectives. Prior work on reactive support from inverter-based DERs and shunt compensation confirms that targeted reactive resources can materially enlarge the admissible operating region \cite{bian2016reactive}, motivating the sensitivity-based capacitor placement strategy developed here.
\subsection{AC XOPF and the Digital Twin Platform}
\begin{figure}[h]
    \centering
    \includegraphics[width=1\linewidth]{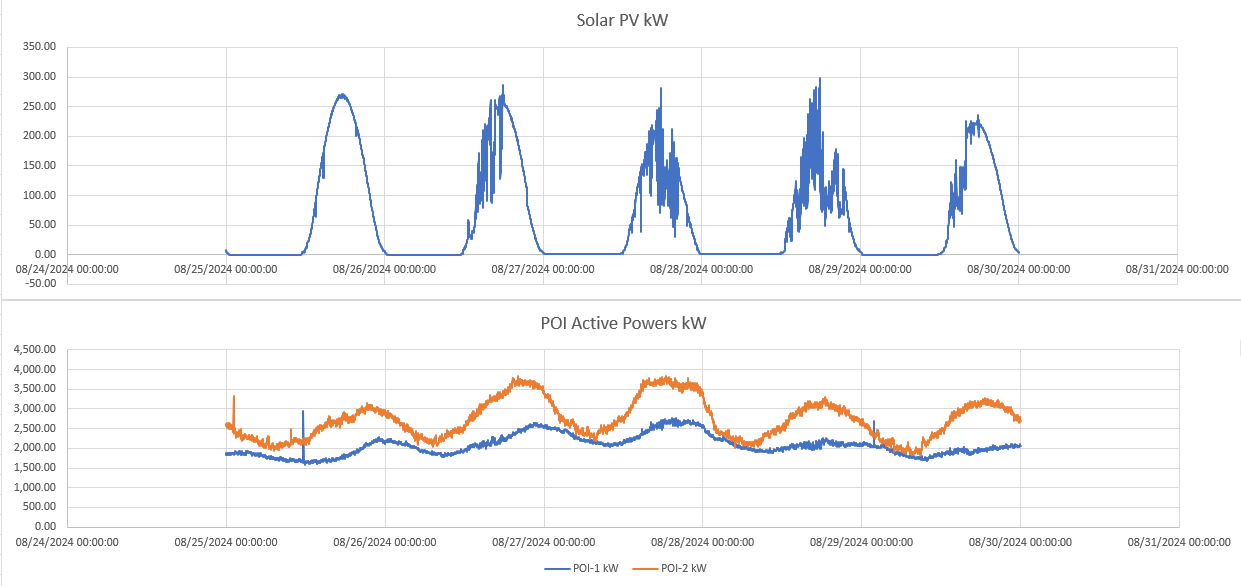}
    \caption{Hourly Demand Recorded over Five Days}
    \label{fig:Load}
\end{figure}

The AC Extended OPF (AC XOPF) provides a unified optimization engine with selectable performance criteria and we utilize both economic dispatch and minimization of load not served. At each solve, the tool returns optimization sensitivities characterizing the marginal impact of reactive-power injection ($\mathrm{OS}_Q$) and voltage relaxation ($\mathrm{OS}_V$) at every bus. These sensitivities drive both corrective scheduling and planning decisions.
The DyMonDS Digital Twin platform \cite{ilic2025dymonds} embeds AC XOPF within a multi-layered simulation environment that models physical power balancing against market incentives. For this study, the platform is configured for time-series analysis of a city-scale microgrid utilizing recorded load profiles from~\ref{fig:Load} and fixed network topology. 
\subsection{System Description}
\begin{figure}[h]
    \centering
    \includegraphics[width=1\linewidth]{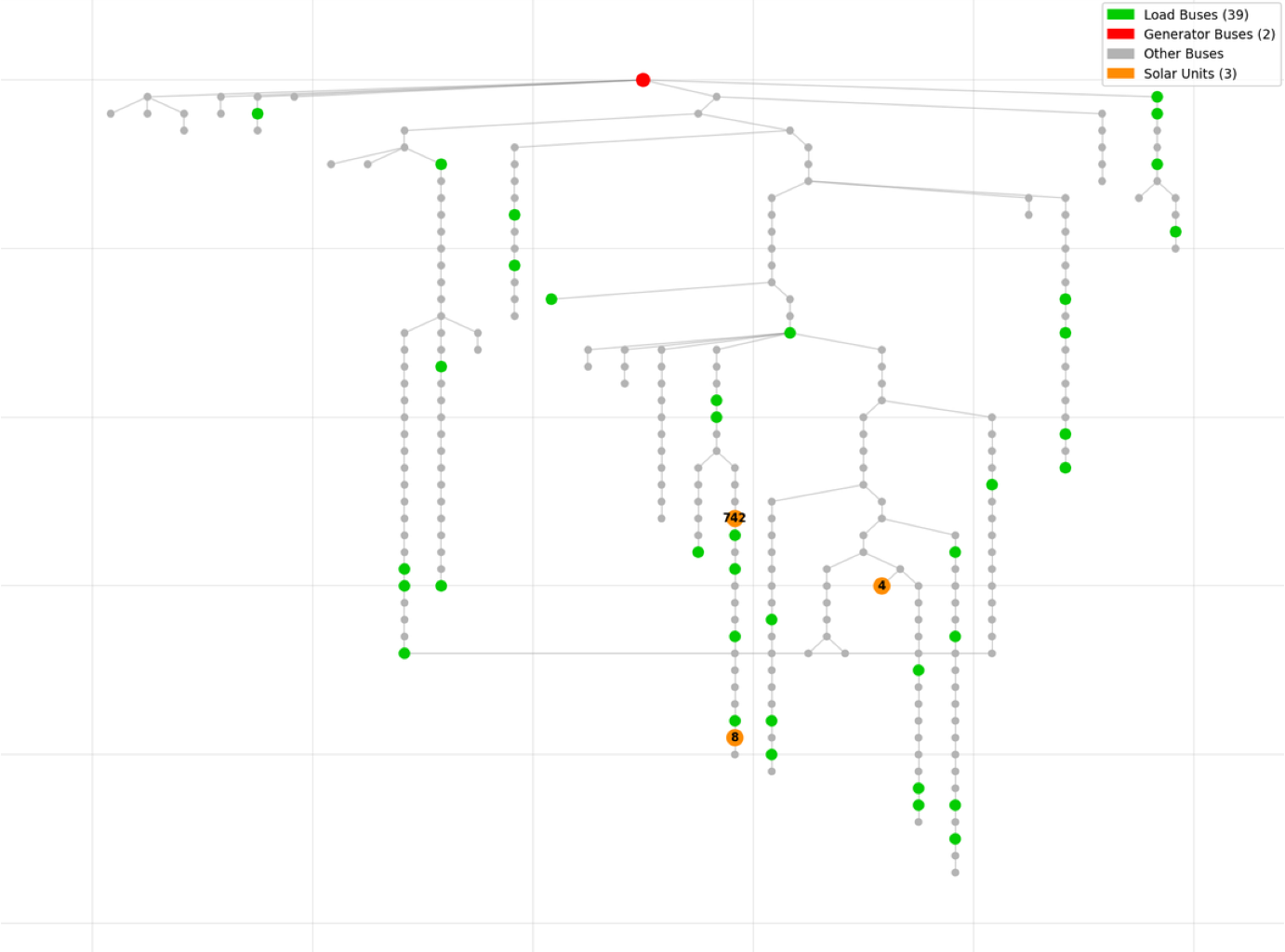}
    \caption{System Topology Plotted from PSS/E Raw File}
    \label{fig:Topology}
\end{figure}
The grid being modeled consists of a small (4 MW Peak) community operating at 12.47 kV. The system has two points of interconnection (POI) with utility power designated as POI-1 and POI-2. For our simulations we fully disconnect the community from utility power by opening both switches connecting POI-1 and POI-2. The microgrid system uses a diesel generator on feeder 1 to support the majority of the load. There is also a solar generation unit directly connected to battery storage on the system that is operating in grid-following mode and modeled as a PV bus. For our simulations, we model three distributed solar units as PQ bus negative loads as shown in Figure~\ref{fig:Topology}.
\subsection{Positioning Relative to Prior Hosting-Capacity Work}
\begin{table*}[t]
\centering
\caption{Positioning of the Present Study Against Related Hosting-Capacity and Microgrid OPF Work}
\label{tab:positioning}
\begin{tabular}{@{}l c c c c c@{}}
\toprule
\textbf{Reference} & \textbf{AC-OPF} & \textbf{Islanded} & \textbf{Multi-Objective} & \textbf{Sensitivity-Guided} & \textbf{Econ.\ Planning} \\
 & & \textbf{MG} & \textbf{Comparison} & \textbf{Cap Placement} & \textbf{(VoLL)} \\
\midrule
Liu et al.\ \cite{liu2023sensitivityHC} & \checkmark & \checkmark & & \checkmark\textsuperscript{a} & \\
Bragin et al.\ \cite{bragin2022mgoperation} & \checkmark & \checkmark & & & \\
Zhang et al.\ \cite{zhang2017robustSCOPF} & \checkmark & & & & \\
Ili\'{c} et al.\ \cite{ilic2024acxopf} & \checkmark & & \checkmark & \checkmark & \\
\textbf{This paper} & \checkmark & \checkmark & \checkmark & \checkmark & \checkmark \\
\bottomrule
\end{tabular}
\end{table*}

Table~\ref{tab:positioning} positions the present study against the most closely related work. The key gap is that no prior study combines: (i) comparative evaluation of multiple OPF objectives on the same islanded microgrid, (ii)a composite sensitivity score for capacitor-placement ranking, and (iii) a planning optimization that frames the decision as investment-versus-VoLL.

\section{Methodology: Four Comparative Cases}
\label{sec:methodology}
At each time step, the digital twin solves AC-OPF with fixed network physics, time-varying demand, and PV injections modeled as negative PQ loads consistent with grid-following inverter control where reactive power is not independently dispatched. The generic formulation minimizes an objective over the set of generators $\mathcal{G}$:
\begin{equation}
\min_{P_G,Q_G,V,\theta} \sum_{i \in \mathcal{G}} C_i(P_{Gi})
\label{eq:obj}
\end{equation}
subject to the nodal AC power-balance equations
\begin{align}
P_{Gi} - P_{Di}
&= \sum_{j=1}^{N} V_i V_j \bigl( G_{ij}\cos(\theta_i - \theta_j)
+ B_{ij}\sin(\theta_i - \theta_j) \bigr), \label{eq:Pbal} \\
Q_{Gi} - Q_{Di}
&= \sum_{j=1}^{N} V_i V_j \bigl( G_{ij}\sin(\theta_i - \theta_j)
- B_{ij}\cos(\theta_i - \theta_j) \bigr), \label{eq:Qbal}
\end{align}
and operating limits
\begin{align}
P_{Gi}^{\min} &\leq P_{Gi} \leq P_{Gi}^{\max}, \label{eq:Plim}\\
Q_{Gi}^{\min} &\leq Q_{Gi} \leq Q_{Gi}^{\max}, \label{eq:Qlim}\\
V_i^{\min} &\leq V_i \leq V_i^{\max}. \label{eq:Vlim}
\end{align}
Feasibility requires the existence of decision variables $\{P_G, Q_G, V, \theta\}$ satisfying all equations and inequalities simultaneously. When no such set exists, the AC-OPF is infeasible. The same high-PV islanded scenario is evaluated through four cases:
\subsection{Case~1 --- Economic Dispatch}
The optimizer minimizes generation cost in islanded mode with a single dispatchable generator, using the priority ordering $\min(\text{cost}) \rightarrow \min(P_g) \rightarrow \min(\text{loss})$. Hard voltage limits are enforced. This case establishes the baseline economic operating point without any corrective actions or hardware modifications.
\subsection{Case~2 --- Voltage Stress (PV Power Factor Variation)}
PV power factors are altered from nominal values, changing the reactive-power contribution of each PV unit and pushing the system toward its voltage-feasibility boundary. This case exposes the sensitivity of the AC-OPF solution to PV reactive-power assumptions: under the same real-power injection profile, changes in inverter power factor can shift the system from comfortably feasible to marginally solvable. The degraded solution quality observed in this case motivates the corrective actions explored in Cases~3 and~4.
\subsection{Case~3 --- Optimal Load Delivery (OLD)}
Load shedding is introduced as a corrective decision variable and minimized. This case addresses the voltage stress identified in Case~2 through operational curtailment rather than infrastructure modification. It quantifies reliability pressure through the location and magnitude of curtailed demand and establishes the cost of maintaining feasibility without capital investment \cite{paul2018demand}.
\subsection{Case~4 --- Capacitor Enhancement (Economic Objective)}
Capacitor upgrades selected via the $S_k$ sensitivity score (Section~\ref{sec:capscore}) are added to the network and the Economic objective from Case~2 is rerun. This case addresses the same voltage stress through preventive infrastructure investment rather than corrective load shedding. It enables an explicit post-processing economic comparison against Case~3: the difference in system cost and load served quantifies the investment justification.

\section{Sensitivity-Guided Capacitor Placement}
\label{sec:capscore}
\subsection{Composite Placement Score}
Capacitor location ranking is based on optimization sensitivities extracted from the comparative runs. For each bus $k$, the placement score combines reactive and voltage sensitivity magnitudes with planner-selected weights:
\begin{equation}
S_k = w_Q\,\bigl|\mathrm{OS}_Q^{(k)}\bigr| + w_V\,\bigl|\mathrm{OS}_V^{(k)}\bigr|
\label{eq:composite}
\end{equation}
where $w_Q + w_V = 1$. Larger $S_k$ indicates higher upgrade priority for capacitor placement. The score $S_k$ admits a direct physical interpretation: $\mathrm{OS}_Q^{(k)}$ captures the marginal reduction in objective per unit of reactive injection at bus $k$, whereas $\mathrm{OS}_V^{(k)}$ captures the marginal impact of relaxing the voltage bound at bus $k$ on total system cost. The convex combination allows planners to bias toward purely reactive-support-driven placement ($w_Q=1$) or toward voltage-limit-driven placement ($w_V=1$) depending on study goals.
\subsection{Mathematical Justification}
The sensitivity quantities $\mathrm{OS}_Q$ and $\mathrm{OS}_V$ are derived from the Karush--Kuhn--Tucker (KKT) conditions of the AC-OPF. Specifically, $\mathrm{OS}_Q^{(k)}$ is associated with the reactive power balance \eqref{eq:Qbal} at bus $k$, and $\mathrm{OS}_V^{(k)}$ relates the optimal cost to changes in the maximum allowable voltage \eqref{eq:Vlim}. At an optimal primal-dual solution, these quantities satisfy
\begin{equation}
\frac{\partial \mathcal{L}}{\partial Q_{Dk}} = \mathrm{OS}_Q^{(k)}, \qquad
\frac{\partial \mathcal{L}}{\partial V_k^{\max}} = \mathrm{OS}_V^{(k)},
\label{eq:kkt_sens}
\end{equation}
where $\mathcal{L}$ is the Lagrangian of the AC-OPF. A shunt capacitor at bus $k$ injecting $\Delta Q_{\mathrm{cap}}$ modifies the reactive balance at bus $k$ and, to first order, reduces the objective by approximately $\mathrm{OS}_Q^{(k)} \cdot \Delta Q_{\mathrm{cap}}$. The score $S_k$ therefore ranks buses by the aggregate benefit of local reactive injection on both cost and voltage margin.

Because the AC-OPF is nonconvex, dual variables are local properties of the computed solution; however, they are consistently informative when the solver returns a KKT point \cite{bolognani2016existence}. The four-case comparison guards against over-reliance on a single objective's sensitivities by exposing where rankings agree (high-confidence placement) and disagree (objective-dependent sensitivity).

\section{Post-Processing Planning Optimization}
\label{sec:planning}
After sensitivity ranking, a planning optimization balances capacitor-upgrade spending against expected reliability penalties. Define the binary decision vector $\mathbf{x} \in \{0,1\}^N$ where $x_k=1$ indicates installation of a capacitor at bus $k$. The planning problem is:
\begin{equation}
\min_{\mathbf{x} \in \{0,1\}^N}
\sum_k x_k\,C_{\mathrm{cap}}(k) + \sum_k (1-x_k)\,C_{\mathrm{VoLL}}(k)
\label{eq:planning}
\end{equation}
where $C_{\mathrm{cap}}(k)$ is the annualized capacitor cost at bus $k$ (procurement, installation, maintenance) and $C_{\mathrm{VoLL}}(k)$ is the expected cost of lost load at bus $k$ over the planning horizon if no capacitor is installed. The expected VoLL term is computed from the Case~3 (OLD) results:
\begin{equation}
C_{\mathrm{VoLL}}(k) = \sum_{t \in \mathcal{T}} \Delta P_{\mathrm{shed},k}(t) \cdot \mathrm{VoLL} \cdot \Delta t
\label{eq:voll}
\end{equation}
where $\Delta P_{\mathrm{shed},k}(t)$ is the load shed at bus $k$ at time $t$ in Case~3 and $\mathrm{VoLL}$ is the per-unit reliability penalty (\$/MWh).

\textbf{Optimality condition.} For bus $k$ with $S_k > 0$, installing a capacitor is optimal when $C_{\mathrm{cap}}(k) < C_{\mathrm{VoLL}}(k)$. The economic comparison of Case~4 versus Case~3 then validates the planning solution: if total system cost under Case~4 (with capacitors) is lower than under Case~3 (with load shedding), the investment is justified.

\section{Results and Discussion}
\label{sec:results}
All four cases were solved over a 47-hour horizon on the islanded microgrid with high PV modeled as negative PQ loads. Every case converged for all 47 valid hours. Table~\ref{tab:results} summarizes the key metrics.

\begin{table*}[t]
\centering
\caption{Summary of Four-Case Simulation Results (47-Hour Horizon)}
\label{tab:results}
\begin{tabular}{@{}l r r r r r r r@{}}
\toprule
\textbf{Case} & \textbf{Total Cost} & \textbf{Load Served} & \textbf{Load Shed} & \textbf{Avg Mismatch} & \textbf{Avg $V_{\min}$} & \textbf{Avg $V_{\max}$} & \textbf{Top Cap Buses} \\
 & \textbf{(\$)} & \textbf{(MW)} & \textbf{(MW)} & & \textbf{(p.u.)} & \textbf{(p.u.)} & \\
\midrule
1 --- Economic    & 8{,}605.26 & 87.81 & ---   & 0.0012 & 1.016 & 1.019 & 508, 364, 675 \\
2 --- Voltage Stress & 5{,}930.71 & 87.02 & ---   & 0.0494 & 1.043 & 1.048 & 508, 364, 675 \\
3 --- OLD         & 7{,}045.99 & 71.74 & 16.07 & 0.0008 & 1.019 & 1.022 & 508, 364, 783 \\
4 --- Cap Enhanced & 8{,}605.34 & 87.81 & ---   & 0.0011 & 1.015 & 1.019 & 508, 364, 675 \\
\bottomrule
\end{tabular}
\end{table*}

\subsection{Case~1: Economic Dispatch Baseline}
\begin{figure}[h]
    \centering
    \includegraphics[width=1\linewidth]{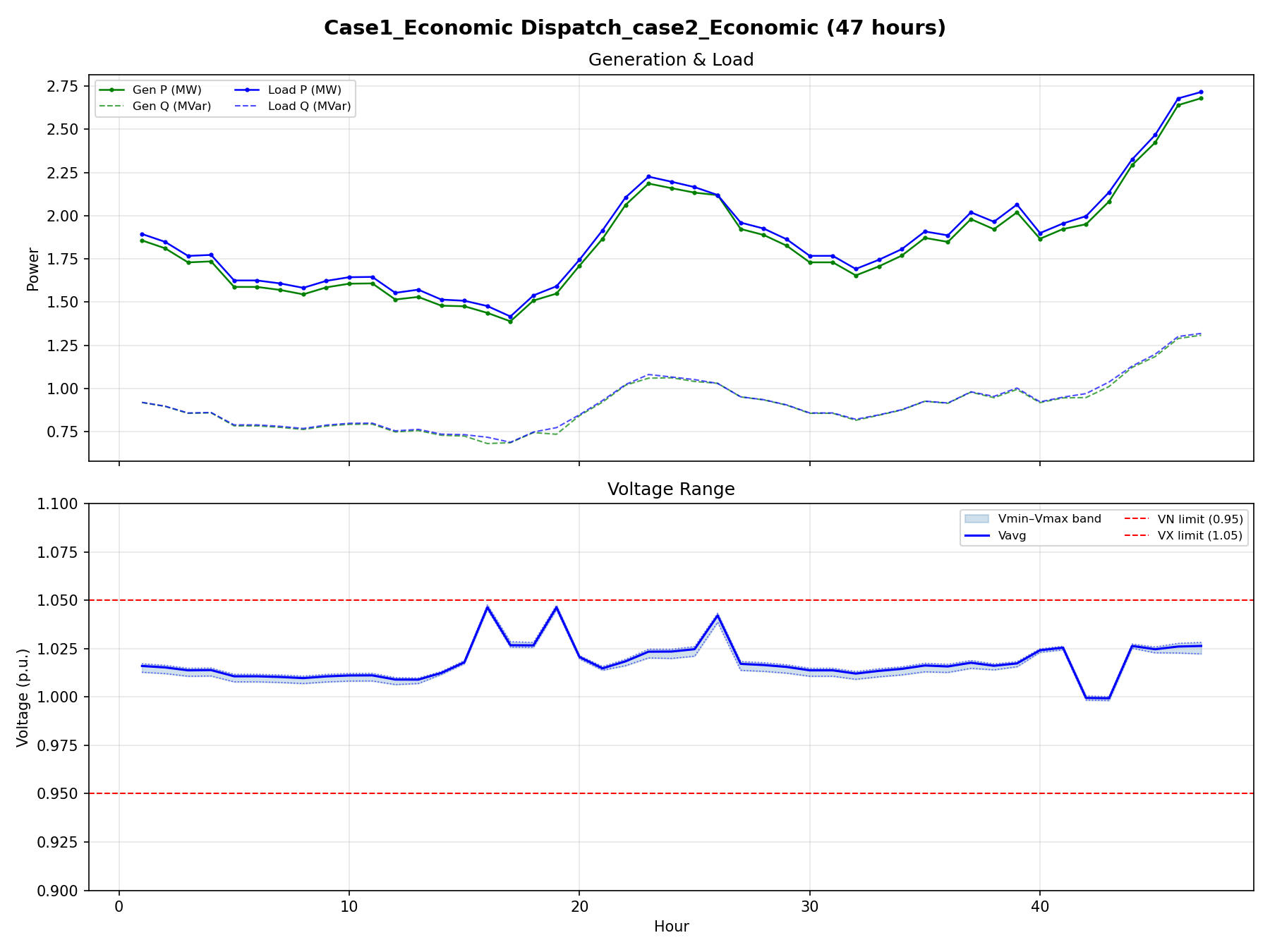}
    \caption{Case 1 --- Time-series dispatch profile: real power, reactive power, losses, and cost.}
    \label{fig:c1_ts}
\end{figure}
The economic dispatch serves the full 87.81~MW load at a total cost of \$8,605 with tight power-balance mismatch (0.0012 average). Bus voltages range from 1.016--1.019 p.u. This case establishes the baseline operating point: the system is economically operable in islanded mode, and the sensitivity analysis identifies buses 508, 364, and 675 as the top capacitor candidates. Figure~\ref{fig:c1_ts} shows the time-series dispatch profile.

\subsection{Case~2: Voltage Stress under PV Power Factor Variation}
\begin{figure}[h]
    \centering
    \includegraphics[width=1\linewidth]{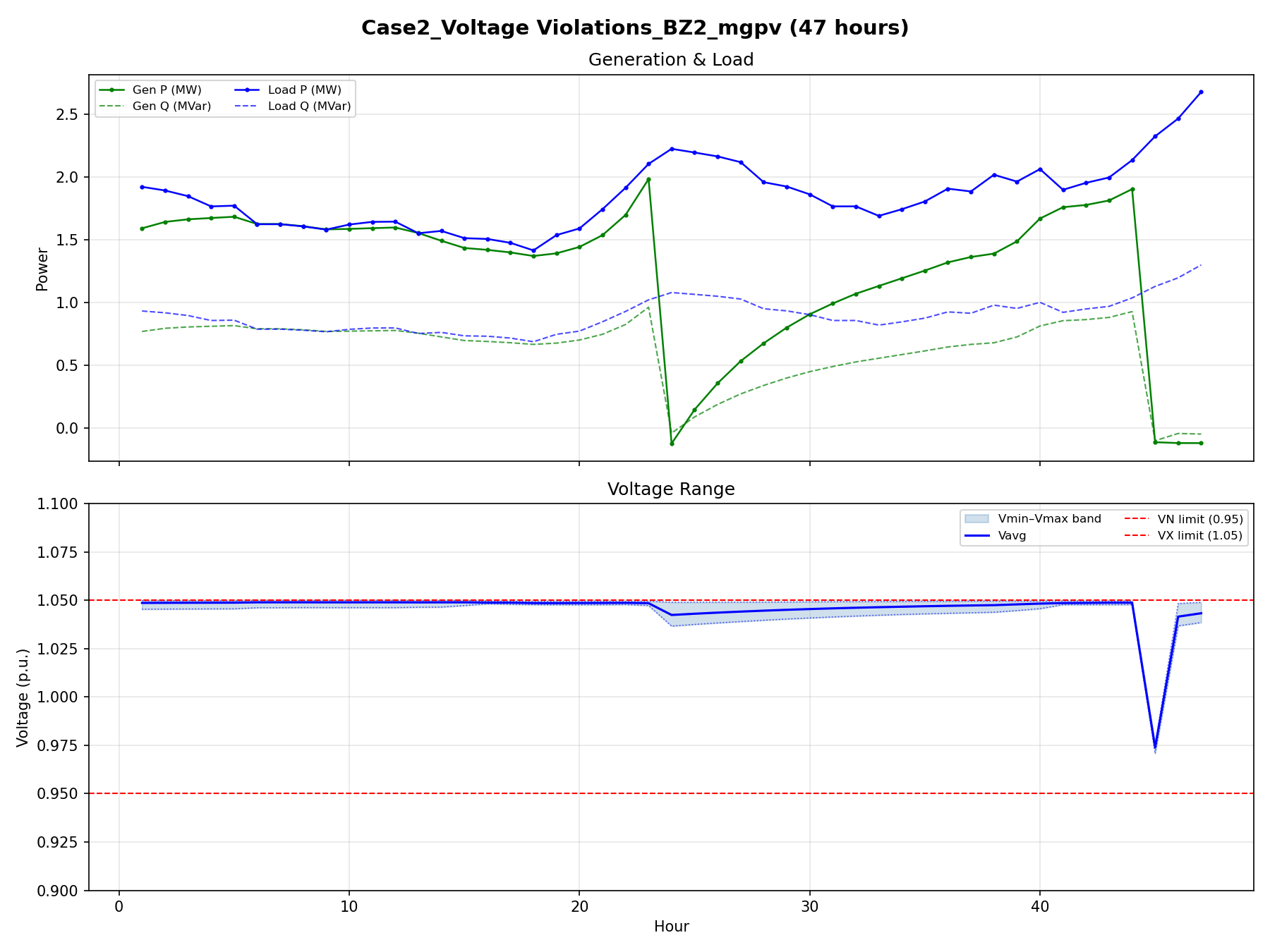}
    \caption{Case 2 --- Time-series dispatch profile under altered PV power factors.}
    \label{fig:c2_ts}
\end{figure}
Lowering PV power factors pushes the system toward its voltage-feasibility boundary. The solution quality degrades substantially: average mismatch increases to 0.0494 (approximately 40 times Case~1), and voltages drift to 1.043--1.05 p.u. Total cost drops to \$5,931 as the solver fails to dispatch at multiple time steps. This case demonstrates that PV reactive-power assumptions are a binding constraint at high penetration---the same real-power injection profile yields a qualitatively different operating regime when inverter power factors change. The same top buses (508, 364, 675) appear as capacitor candidates, indicating that the stress pattern is geographically consistent with Case~1. Figure~\ref{fig:c2_ts} shows time-series results, then cases~3 and~4 address the degraded operating conditions exposed here through two contrasting corrective strategies.

\subsection{Case~3: Optimal Load Delivery (Corrective Load Shedding)}
\begin{figure}[h]
    \centering
    \includegraphics[width=1\linewidth]{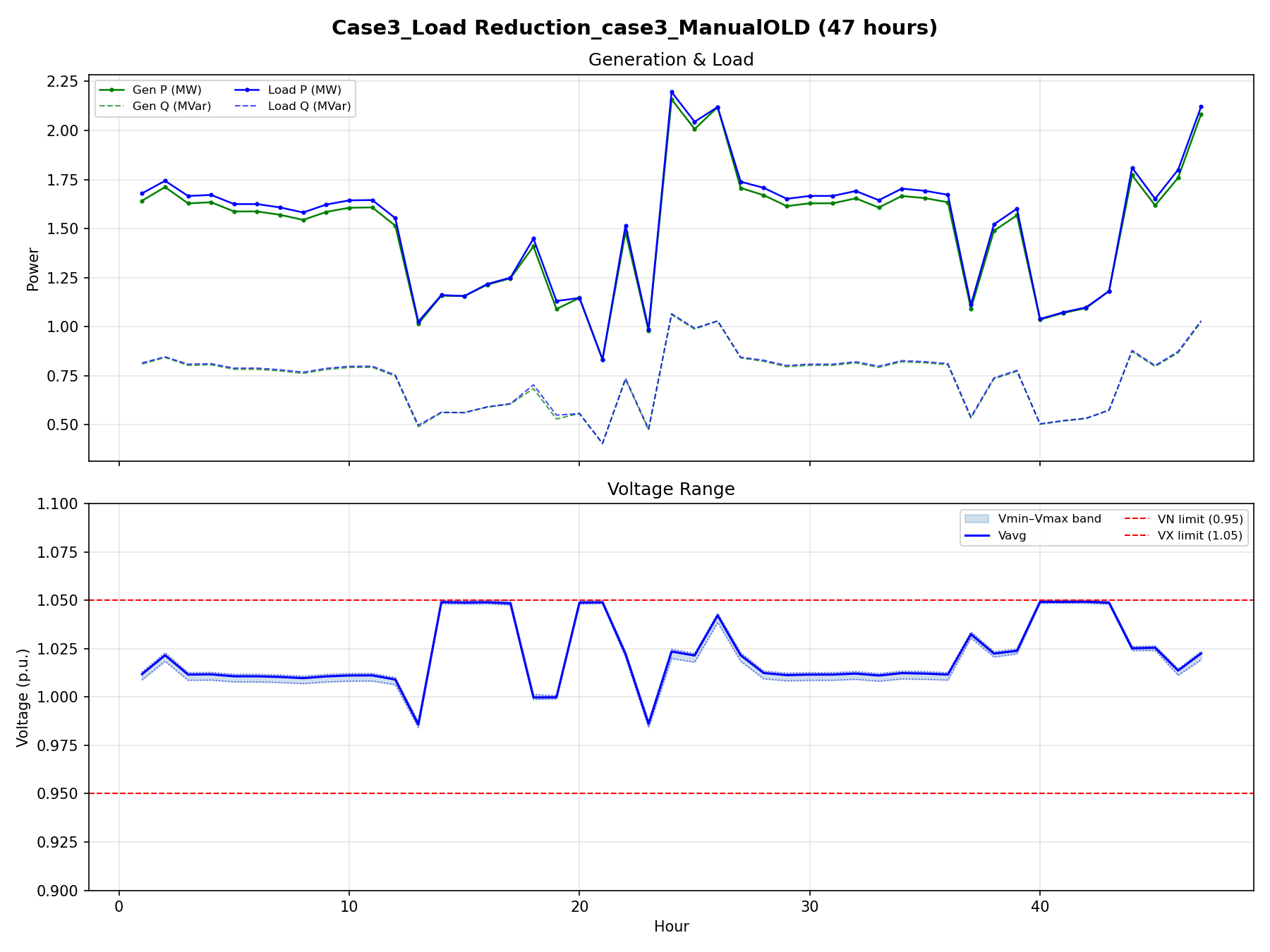}
    \caption{Case 3 --- Time-series dispatch profile with corrective load shedding.}
    \label{fig:c3_ts}
\end{figure}
The OLD formulation addresses the voltage stress identified in Case~2 by introducing voluntary load shedding as a corrective action. The optimizer sheds a total of 16.07~MW across the horizon, reducing served load from 87.81 to 71.74~MW at a total cost of \$7,046. Power-balance mismatch drops to the lowest of all cases (0.0008 average), and voltages narrow to 1.019--1.022 p.u., confirming that curtailment restores high-quality feasibility. The third-ranked capacitor bus shifts from 675 to 783, indicating that load redistribution from shedding changes the reactive stress pattern and therefore the optimal placement recommendation. This case quantifies the cost of maintaining feasibility without infrastructure investment: 16.07~MW of demand must be curtailed. Figure~\ref{fig:c3_ts} shows the dispatch results.

\subsection{Case~4: Capacitor-Enhanced Economic Dispatch}

\begin{figure}[h]
    \centering
    \includegraphics[width=1\linewidth]{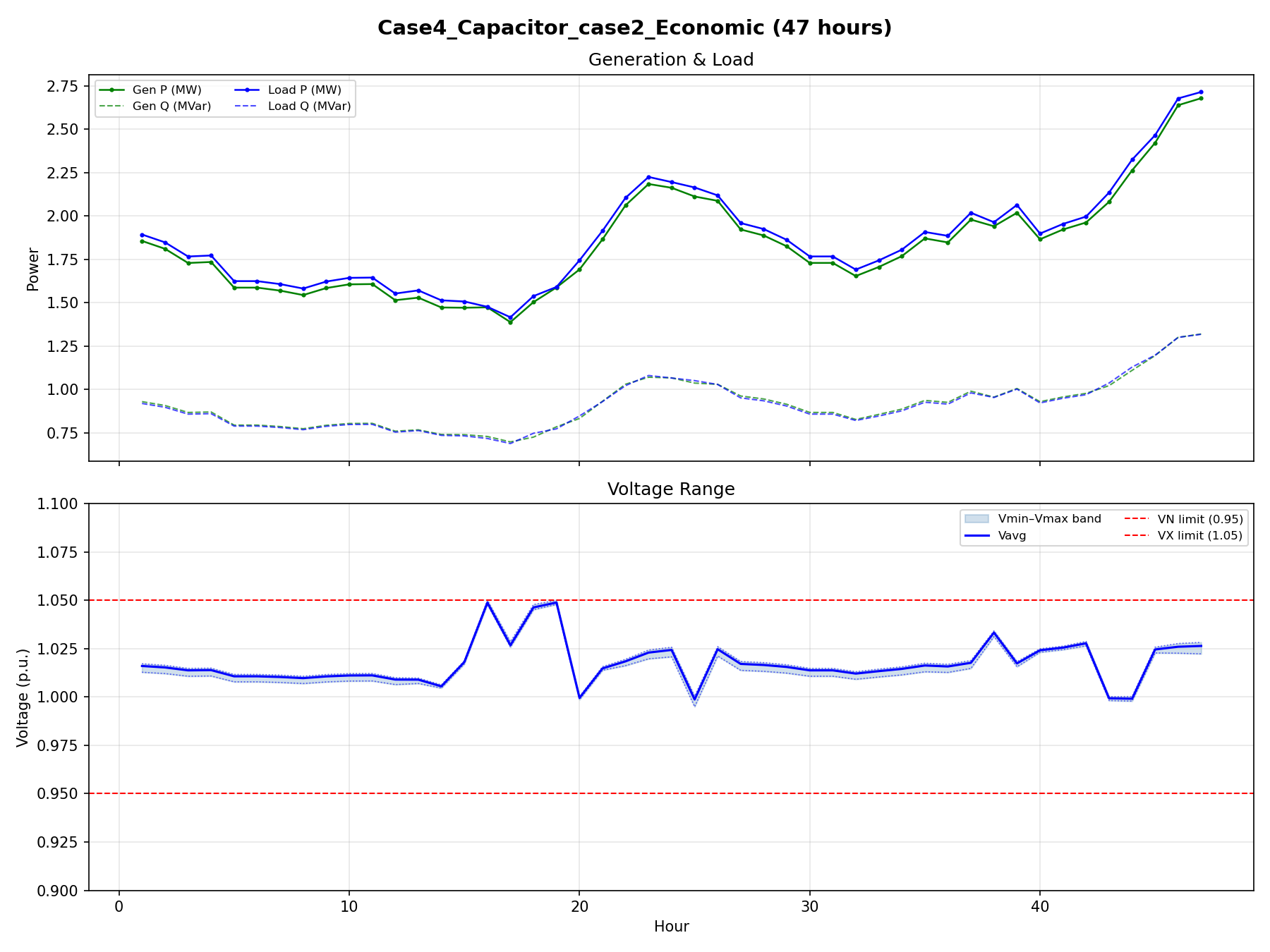}
    \caption{Case 4 --- Time-series dispatch profile with capacitors at buses 508, 364, 675.}
    \label{fig:c4_ts}
\end{figure}

\begin{figure}[h]
    \centering
    \includegraphics[width=1\linewidth]{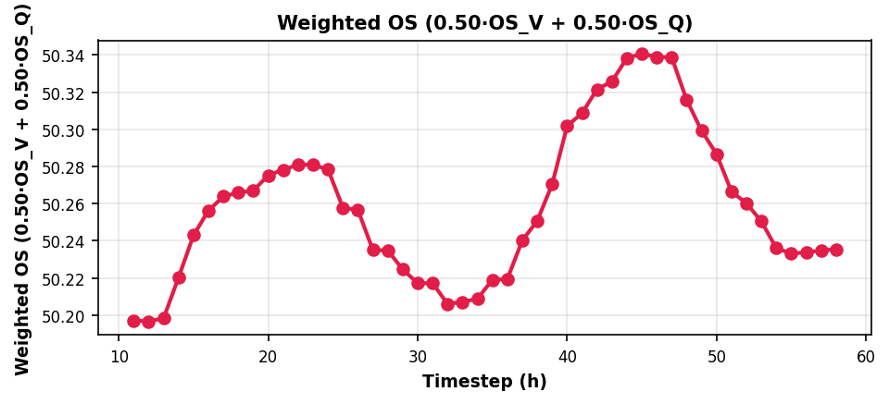}
    \caption{Weighted sensitivity score $S_k = 0.5\,|\mathrm{OS}_V| + 0.5\,|\mathrm{OS}_Q|$ used for capacitor placement ranking.}
    \label{fig:c4_os}
\end{figure}

Capacitors are placed at buses 508, 364, and 675---the top-ranked locations from the $S_k$ composite score---and the Economic objective from Case~1 is rerun. Full load (87.81~MW) is served at a total cost of \$8,605.34, essentially identical to Case~1. Average mismatch is 0.0011, and voltages tighten to 1.015--1.019~p.u. The capacitor placement restores the full load service that Case~3 achieved through curtailment, with no measurable cost penalty relative to the economic baseline. Figure~\ref{fig:c4_ts} displays the time-series results, and figure~\ref{fig:c4_os} shows the weighted sensitivity score $S_k$ used for placement.
\subsection{Economic Comparison: Case~4 vs Case~3}
\begin{figure}[h]
    \centering
    \includegraphics[width=1\linewidth]{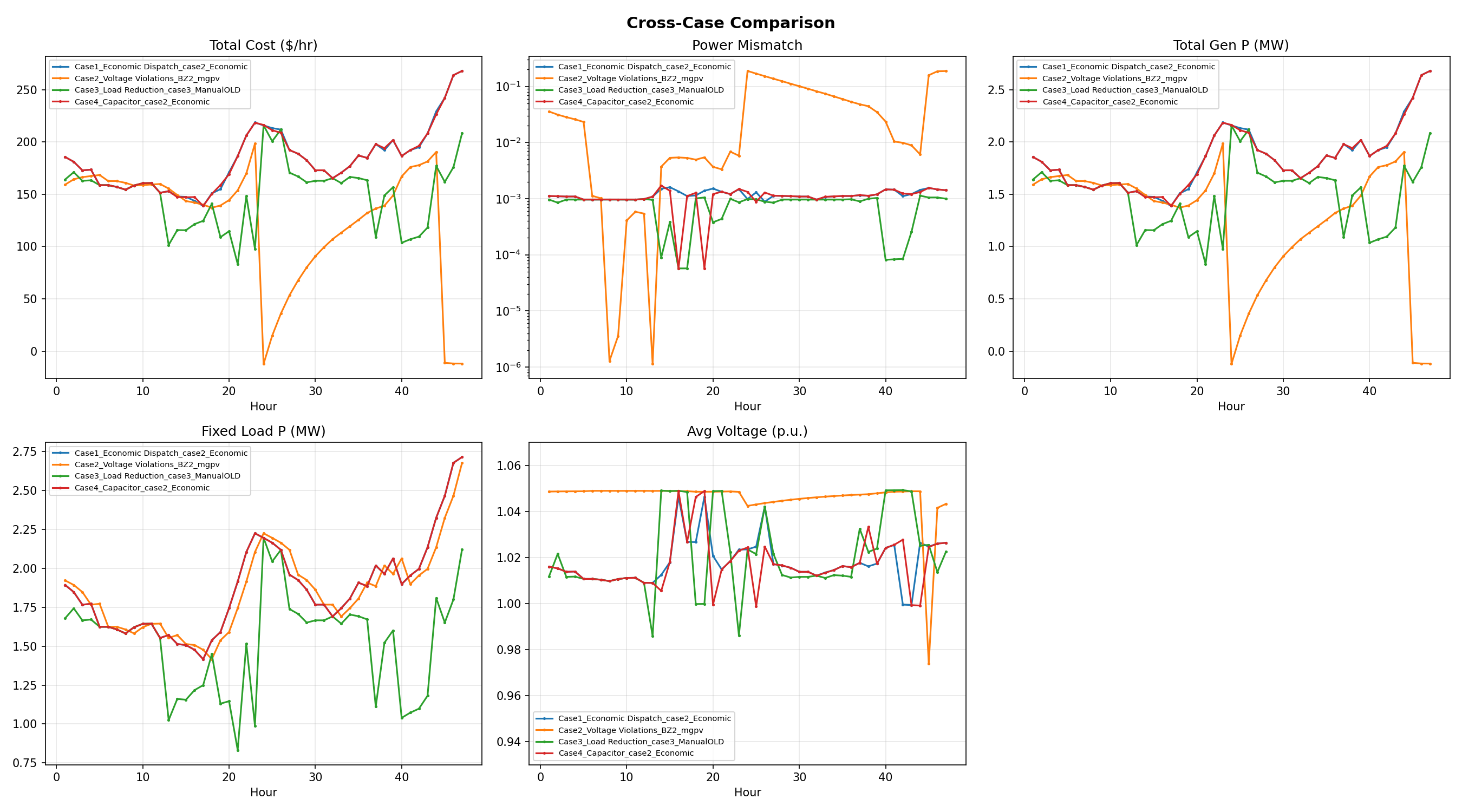}
    \caption{Cross-case comparison of total cost, load served, mismatch, and voltage range across all four cases.}
    \label{fig:cross_case}
\end{figure}
The planning trade-off between infrastructure investment and corrective curtailment is quantified by comparing Cases~3 and~4 directly. Capacitor enhancement recovers 16.07~MW of curtailed load at an additional operating cost of \$1,559 over the 47-hour horizon, corresponding to \$97 per MW of recovered demand. This value provides the threshold for the planning decision from \eqref{eq:planning}: if the site-specific VoLL exceeds \$97/MW, capacitor installation is cost-justified. Both cases correct the voltage stress exposed in Case~2, but through opposite mechanisms---corrective curtailment versus preventive infrastructure---giving the planner a direct, data-driven input to the investment-versus-reliability trade-off. Figure~\ref{fig:cross_case} presents the cross-case comparison of key metrics.

\subsection{Sensitivity Ranking Consistency}
Buses 508 and 364 rank first and second across all four cases, indicating high-confidence placement candidates regardless of objective formulation. The third-ranked bus differs between cases with load shedding (bus 783 in Case~3) and cases without (bus 675 in Cases~1, 2, and 4). This confirms that the four-case comparison provides information that a single-objective study would miss: the load-redistribution effect of corrective shedding shifts the reactive stress pattern and alters the marginal placement recommendation.

\section{Conclusion}
\label{sec:conclusion}
This study demonstrates that high-PV islanded microgrid operation is limited by constrained nonlinear feasibility, not only by bulk real-power balance. Time-series AC-OPF in a digital twin provides a practical mechanism to detect when feasible operating sets collapse and to identify whether the driver is reactive-voltage stress or active-power scarcity. The four-case comparative methodology reveals that different objective formulations expose different constraint-binding patterns, producing complementary planning signals that a single-objective study would miss.
Sensitivity-guided capacitor placement provides a direct bridge from operations to planning: the composite score $S_k$ ranks candidate buses using optimization-derived quantities with clear physical interpretation, and the post-processing comparison frames the decision in standard investment-versus-reliability terms. The economic comparison of Case~4 (cap-enhanced) versus Case~3 (load-shed) closes the loop by quantifying the return on infrastructure investment at \$97/MW of recovered demand.
The methodology generalizes beyond the specific network studied. The recovery options, sensitivity-based ranking, and planning optimization are network-agnostic; only the numerical results are topology- and profile-dependent. Future work includes battery storage as an alternative or complement to capacitors and dynamic calculations for real-time decision making.
\section*{Acknowledgment}
Use of the SMartGridz, Inc.\ AC OPF software is greatly appreciated. The specific operational options employed in Cases~1 through~4 are described in \cite{ilic2020derms}.

\bibliographystyle{IEEEtran}
\bibliography{references}

\end{document}